\documentclass{llncs}
\pdfoutput=1
\usepackage[linesnumbered,ruled]{algorithm2e}

\usepackage{graphicx}
\usepackage[tight,footnotesize]{subfigure}

\usepackage{cite}

\usepackage{todonotes}
\newcommand{\xkaapi}{XKaapi}

\newcommand{\dpotrf}{\texttt{DPOTRF}}
\newcommand{\dgetrf}{\texttt{DGETRF}}
\newcommand{\dgeqrf}{\texttt{DGEQRF}}

\begin{document}
\title{
	Scheduling data flow program in \xkaapi: \\
	A new affinity based Algorithm for Heterogeneous Architectures
}
%
%
\author{
	Rapha\"{e}l Bleuse\inst{1} \and
	Thierry Gautier\inst{2} \and
	Jo\~{a}o V. F. Lima\inst{4} \and
	Gr\'{e}gory Mouni\'{e}\inst{1} \and
	Denis Trystram\inst{1}\inst{3}
}
%
%
\tocauthor{
	Rapha\"{e}l Bleuse,
	Thierry Gautier,
	Jo\~{a}o V. F. Lima,
	Gr\'{e}gory Mouni\'{e} and
	Denis Trystram
}
\institute{
	Univ. Grenoble Alpes, France
\and
	Inria Rh\^{o}ne-Alpes, France
\and
	Institut universitaire de France
\and
	Federal University of Rio Grande do Sul (UFRGS), Porto Alegre, Brazil
}

\maketitle              

\begin{abstract}
Efficient implementations of parallel applications on heterogeneous hybrid
architectures require a careful balance between computations and communications
with accelerator devices.
Even if most of the communication time can be overlapped by computations, it is
essential to reduce the total volume of communicated data.
The literature therefore abounds with {\em ad hoc} methods to reach that
balance, but that are architecture and application dependent.
We propose here a generic mechanism to automatically optimize the scheduling
between CPUs and GPUs, and compare two strategies within this mechanism: the
classical Heterogeneous Earliest Finish Time (HEFT) algorithm and our new,
parametrized, Distributed Affinity Dual Approximation algorithm (DADA), which
consists in grouping the tasks by affinity before running a fast dual
approximation.
We ran experiments on a heterogeneous parallel machine with six CPU cores and
eight NVIDIA Fermi GPUs.
Three standard dense linear algebra kernels from the PLASMA library have been
ported on top of the \xkaapi{} runtime.
We report their performances.
It results that HEFT and DADA perform well for various experimental conditions,
but that DADA performs better for larger systems and number of GPUs, and, in
most cases, generates much lower data transfers than HEFT to achieve the same
performance.
\keywords{heterogeneous architectures, scheduling, cost models, dual
approximation scheme, programming tools, affinity}
\end{abstract}
%
\section{Introduction}
%
With the recent evolution of processor design, the future generations of
processors will contain hundreds of cores. To increase the performance per watt
ratio, the cores will be non-symmetric with few highly powerful cores (CPU) and
numerous, but simpler, cores (GPU). The success of such machines will rely on the
ability to schedule the workload at runtime, even for small problem instances.

One of the main challenges is to define a scheduling strategy that may be able
to exploit all potential parallelisms on a heterogeneous architecture composed
of multiple CPUs and multiple GPUs.
Previous works demonstrate the efficiency of strategies such as static
distribution~\cite{SongDongarra2012, TomovDongarraBaboulin2010},
centralized list scheduling with
data locality~\cite{BuenoPlanasDuranEtAl2012}, cost models
\cite{AgulloAugonnetDongarraEtAl2011a,
AgulloAugonnetDongarraEtAl2011,AugonnetThibaultNamyst2010,
AugonnetThibaultNamystEtAl2011} based on Heterogeneous-Earliest-Finish-Time scheduling (HEFT)
\cite{TopcuogluHaririWu2002}, and dynamic for some specific application domains
\cite{BosilcaBouteillerDanalisEtAl2012, HermannRaffinFaureEtAl2010}.
Locality-aware work stealing~\cite{GautierLimaMaillardEtAl2013}, with a careful
implementation to overlap communication by computation~\cite{LimaGautierMaillardEtAl2012}, improves significantly the
performance of compute-bound linear algebra problems such as matrix product and Cholesky factorization.

Nevertheless, none of the above cited works considers scheduling strategies from
the viewpoint of a compromise between performance and locality.  In this paper,
we propose a scheduling algorithm based on dual approximation~\cite{Kedad-SidhoumMonnaMounieEtAl2013} that uses a performance model to
predict the execution time of tasks during scheduling decision.
This algorithm, called Distributed Affinity Dual Approximation (DADA), is able to find a compromise between transfers and performance.
It is parametrized by $\alpha$ for tuning this trade-off.
The main advantage of dual approximation algorithms is their theoretical performance guarantee as they have a constant approximation ratio.
On the contrary, the worst case of HEFT can be arbitrarily bad~\cite{Kedad-SidhoumMonnaMounieEtAl2013}.

We compare these two different scheduling strategies for data-flow
task programming. These strategies are implemented on top of the
\xkaapi{} scheduling framework with performance models for task
execution time and transfer prediction.  The contributions of this
paper are first the design and implementation of dual approximation
scheduling algorithms (with and without affinity) and second its
evaluation in comparison to the well-known HEFT algorithm on three
dense linear algebra algorithms in double precision floating-point
operations from PLASMA~\cite{ButtariLangouKurzakEtAl2009}: namely
Cholesky, LU, and QR.
To our knowledge, this paper is the first report of experimental
evaluations studying the impact of data transfer model and contention on a
machine with up to $8$ GPUs.

The main lesson of this work is that scheduling algorithms need extra information in order to take
the right decisions.
Such extra information could be obtained in a precise communication model to predict
processing time of each task or in a more flexible information such as the affinity in DADA. 
Even if HEFT remains a good candidate for scheduling such linear algebra kernels, DADA is highly competitive against it
for multi-GPU systems: the experimental results demonstrate that it 
achieves the same range of performances while reducing significantly the communication volume.


The remainder of this paper is organized as follows.
Section~\ref{sec:intro:xkaapi} provides an overview of \xkaapi{} runtime
system, describes the \xkaapi{} scheduling
framework and  the cost model applied for
performance prediction. Section~\ref{sec:sched:strategies} details the two studied
scheduling strategies. Section~\ref{sec:results} presents our
experimental results on a heterogeneous architecture composed of $12$ CPUs and
$8$ GPUs. In Section~\ref{sec:related} we briefly survey related works on
runtime systems, scheduling strategies and performance prediction.  
Finally, Section~\ref{sec:conclusion} concludes the paper and suggests future directions.
%
\section{Scheduling framework in \xkaapi}
\label{sec:intro:xkaapi}
%
The \xkaapi\footnote{http://kaapi.gforge.inria.fr} data-flow model
\cite{GautierBesseronPigeon2007} -- as in Cilk, Intel TBB, OpenMP-3.0,
or OmpSs \cite{BuenoPlanasDuranEtAl2012} -- enables non-blocking task creation:
the caller creates the task and proceeds with the program execution.
Parallelism is explicit while the detection of synchronizations is implicit
\cite{GautierBesseronPigeon2007}: dependencies between tasks and memory
transfers are automatically managed by the runtime.  

\xkaapi{} runtime is structured around the notion of \emph{worker}:
it is the internal representation of kernel thread. 
It executes the code of the tasks takes local scheduling decisions. 
Each \emph{worker} owns a local queue of ready tasks. Our interface is mainly inspired by work stealing scheduler and
composed of three operations that act on workers' queues of tasks: \emph{pop},
\emph{push} and \emph{steal}.  In our previous work, we demonstrated the
efficiency of \xkaapi{} locality-aware work stealing as well as the
corresponding multi-GPU runtime support \cite{GautierLimaMaillardEtAl2013}
using specialized implementation of these operations. A new operation, called \emph{activate},
has been defined to push ready task to a worker's queue.
%
%
%
\subsection{Execution flow}
The sketch of the execution mechanism is the following: at each step, either the own queue of worker is
not empty and the worker uses it; or the worker emits a steal request to a
randomly selected worker in order to get a task to execute.
%
%
According to the dependencies between tasks, once a worker performs a task, it calls the
\emph{activate} operations in order to activate the successors of the task
which become ready for execution.

The \xkaapi{} runtime gets information from each internal events (such as start-end of task execution, or start-end of communication to GPU)
to calibrate the performance model and corrects erroneous predictions due to
unpredictable or unknown behavior ({\em e.g.} operating system state or I/O disturbance). 
StarPU~\cite{ AugonnetThibaultNamystEtAl2011} uses similar runtime measurements in order to correct the performance predictions in their HEFT implementation.

%
%

All of our scheduling strategies follow this sketch.
Every worker terminates its execution when all the tasks of the application have been executed.
\subsection{Pop, Push, Steal and Activate Operations}
A framework interface for scheduling strategies is not a new concept in
heterogeneous systems. Bueno~{\em et al.}~\cite{BuenoPlanasDuranEtAl2012} and
Augonnet~{\em et al.}~\cite{AugonnetThibaultNamystEtAl2011} described a minimal
interface to design scheduling strategies with selection at runtime. 
However, there is little information available on the comparison of different
strategies. Most of them reported performance on centralized list scheduling
and performance models. Our framework is composed of three classical operations in work stealing context, 
plus an action to activate tasks when predecessors have completed.
\begin{itemize}
\item The \emph{push} operation inserts a task into a queue. A worker can push a task into any other workers' queue.
\item A \emph{pop} removes a task from the local queue owned by the caller worker. 
%
\item A \emph{steal} removes a task from the queue of a remote worker. 
It is called by an idle thread -- the \emph{thief} -- in
order to pick tasks from a randomly selected worker -- the \emph{victim}. 
%
%
%
\item The \emph{activate} operation is called after the completion of a task.
The role of this operation is to allocate the tasks that are ready to be executed.
Hence, most of the scheduling decision are done during this operation.
\end{itemize}
%
\subsection{Performance Model}
\label{sec:perf:model}
%
Cost models depend on a certain knowledge of both application algorithm and the
underlying architecture to predict performance at runtime.  In order to predict
performance, we designed a StarPU~\cite{AugonnetThibaultNamyst2010} like
performance model for task execution time  and communication. Our task
prediction relies on an history-based model, and transfer time estimation is
based on asymptotic bandwidth. They are associated with scheduling strategies
that are based on task completion time such as HEFT and DADA with and without affinity.

In order to balance efficiently the load, for each processor \xkaapi{} maintains a shared
time-stamp of the predicted date when it has completed its tasks. 
The completion date of the last executed task is also kept.
The update and incrementation of the time-stamps are efficiently implemented with atomic operators.
\section{Scheduling Strategies}
\label{sec:sched:strategies}
%
This section introduces the scheduling strategies designed on top of the \xkaapi{} scheduling framework. 
We consider a multi-core parallel architecture with $m$ homogeneous CPUs and $k$ homogeneous GPUs. First, we
describe our implementation of HEFT~\cite{TopcuogluHaririWu2002}.
Then, we recall the principle of the dual approximation scheme~\cite{HochbaumShmoys1987}. 
We propose a new algorithm -- Distributed Affinity Dual Approximation (DADA) -- based on this paradigm which takes into account the affinity between tasks.

In the following we denote by $p^{CPU}_i$ the processing time of
task $T_i$ on a CPU and $p^{GPU}_i$ on a GPU.  We define the speedup $S_i$ of
task $T_i$ as the ratio $S_i = p^{CPU}_i\!/\,p^{GPU}_i$.
%
%
\subsection{HEFT within \xkaapi}
The Heterogeneous Earliest-Finish-Time algorithm (HEFT), proposed by
\cite{TopcuogluHaririWu2002}, is a scheduling algorithm for a bounded number of heterogeneous processors.  
Its time complexity is in $O(n^{2} \cdot (m+k))$. 
It has two major phases: \emph{task prioritizing} and a \emph{worker selection}.
Our \xkaapi{} version of HEFT implements both phases during the \emph{activate} operation.
The \emph{task prioritizing} phase computes for all ready tasks $T_i$ its speedup $S_i$ relative to an execution on GPU. 
Next, it sorts the list of ready tasks by decreasing speedups.
Whereas the original HEFT rule sorts the tasks by decreasing upward rank (average path length to the end),
our rule gives priority on minimizing the sum of the execution times.
In the \emph{worker selection} phase, the algorithm selects tasks in the order
of their speedup $S_i$ and schedules each task on the worker which minimizes the completion time. 
Algorithm~\ref{sec:algo:heft} describes the basic steps of HEFT over \xkaapi.

\begin{algorithm}[tb]
\DontPrintSemicolon
\SetKwInOut{Input}{Input}\SetKwInOut{Output}{Output}
\SetKwData{List}{queue}
\SetKwData{Lists}{queues}
\SetKwData{LR}{LR}
\SetKwData{Up}{upper}
\SetKwData{Low}{lower}
\SetKwData{Td}{T}
\SetKwData{activate}{activate}
\SetKwData{lpush}{local\_push}
\SetKwData{rpush}{push}
\SetKwData{pop}{pop}

\Input{A list of ready tasks $T_i$ \LR}
\Output{Tasks $T_i$ pushed to the worker's \Lists}
\BlankLine

\ForEach{$T_i \in \LR$ }{
	$S_i \leftarrow p^{CPU}_i\!/\,p^{GPU}_i$\;
}
Sort all ready tasks $T_i$ by decreasing speedup $S_i$\;
\ForEach{$T_i \in \LR$ }{
	Schedule $T_i$ on the worker $w_j$ achieving the earliest finish time\; \label{sec:algo:heft:sched}
	\rpush of $T_i$ into \List of worker $w_j$\;
	Update processor load time-stamps on worker $w_j$\;
}
\caption{HEFT -- \emph{activate} operation.}
\label{sec:algo:heft}
\end{algorithm}

\subsection{Dual Approximation and Affinity}
\subsubsection{Dual Approximation} Let us recall first that a $\rho$-dual approximation scheduling algorithm
considers a {\it guess} $\lambda$ (which is an estimation of the optimal
makespan) and either delivers a schedule of makespan at most $\rho \lambda$ or
answers correctly that there exists no schedule of length at most $\lambda$
\cite{HochbaumShmoys1987}. The process is repeated by a classical binary search
on $\lambda$ up to a precision of $\epsilon$.
We target $\rho = 2$.
The dual approximation part of Algorithm~\ref{sec:algo:dda} consists in the following steps:
\begin{itemize}
\item Choice of the initial guess $\lambda$ (lines \ref{sec:algo:dda:step1} and
	\ref{sec:algo:dda:step2});
\item Extract the tasks which fit only into GPUs ($p^{CPU}_i>\lambda$), and
	symmetrically those which are dedicated to CPUs (line
	\ref{sec:algo:dda:step3});
\item Keep this schedule if the tasks fit into $\lambda$ (line
	\ref{sec:algo:dda:step4}). Otherwise, reject it if there is
	a task larger than $\lambda$ on both CPUs and GPUs (line
	\ref{sec:algo:dda:step5});
\item Add to the tasks allocated to the GPU those which have the largest
	speedup $S_i$ up to overreaching the threshold $\lambda$ (line
	\ref{sec:algo:dda:step6}) which guarantees the ratio $\rho=2$;
\item Put all the remaining tasks in the $m$ CPUs and execute them using an
	earliest-finish-time scheduling policy (line \ref{sec:algo:dda:step6}).
\end{itemize}
\subsubsection{Affinity} \label{sec:affinity}
DADA builds a compromise taking into account both raw performance and transfers.
The principle consists in two successive phases: a first local phase targeting
the reduction of the communications through the abstraction described below and a second phase which counter-balances the induced serialization
aiming at a global balance. Any algorithm optimizing the makespan could be used
for the second phase. We use a basic dual-approximation. 
In order to gain a finer control, the length of the first phase is controlled
by a parameter (denoted by $\alpha$, $0 \le \alpha \le 1$). A value of $0$ for
$\alpha$ means that the affinity is not taken into account: DADA is then
a basic dual-approximation. While at the opposite a value close to $1$ allows a
length up to $\lambda$ for the first phase, thus giving a greater weight to affinity.

Each pair (task, computation resource) is given an affinity score.
Maximizing the score over the whole schedule enables to consider local impacts. 
The affinity scores are computed using extra information of the runtime.
In our implementation, they were computed using the amount of data updated by each task. 
For instance, a task that \emph{writes} or \emph{modifies} a data stored on a resource $R$ has a high score and is prone to be scheduled on $R$.
\begin{algorithm}[tb]
\DontPrintSemicolon
\SetKwInOut{Input}{Input}\SetKwInOut{Output}{Output}
\SetKwData{List}{queue}
\SetKwData{Lists}{queues}
\SetKwData{LR}{LR}
\SetKwData{Up}{upper}
\SetKwData{Low}{lower}
\Input{A list of ready tasks $T_i$ \LR}
\Output{Tasks $T_i$ pushed to the worker's \Lists}
\BlankLine
$lower \leftarrow 0$\;
$upper \leftarrow \sum_i max(p^{CPU}_i, p^{GPU}_i)$\;\label{sec:algo:dda:step1}
\While{$(upper - lower) > \epsilon$}{
	$\lambda \leftarrow (upper + lower) /\,2$\;\label{sec:algo:dda:step2}

	\Begin(\emph{local affinity phase})
	{
        Schedule tasks of \LR per affinity score on its affinity processor, loading each processor up to overreaching  $\alpha \lambda$
	}

	\Begin(\emph{global balance phase}){
		Schedule \LR to minimize finish time using $\lambda$ as hint\;\label{sec:algo:dda:step3}
		\eIf{tasks do fit into $(2 + \alpha) \lambda$}{
			$upper \leftarrow \lambda$\;
			Keep current schedule\;\label{sec:algo:dda:step4}
		}{
			$lower \leftarrow \lambda$\;
			Reject current schedule\;\label{sec:algo:dda:step5}
		}
	}
}
Push each task $T_i$ of \LR on queue of worker $w_j$ based on the last fitting schedule and update processor load time-stamps\;\label{sec:algo:dda:step6}
\caption{DADA -- \emph{activate} operation.}
\label{sec:algo:dda}
\end{algorithm}
%
\section{Experiments}
\label{sec:results}
%
%
\subsection{Experimental setup: Platform \& Benchmarks}
\label{sec:subsec:perf:setup}
\subsubsection{Platform}
All experiments have been conducted on a heterogeneous, multi-GPU system.  It
is composed of two hexa-core Intel Xeon X5650  CPUs running at 2.66~GHz with
72~GB of memory. It is enhanced with eight NVIDIA Tesla C2050 GPUs (Fermi
architecture) of 448 GPU cores (scalar processors) running at 1.15~GHz each
(2688 GPU cores total) with 3~GB GDDR5 per GPU (18~GB total).
The machine has $4$~PCIe switches to support up to $8$~GPUs.
When $2$~GPUs share a switch, their aggregated PCIe bandwidth is bounded by
the one of a single PCIe~16x.
Experiments using up to $4$~GPUs avoid this bandwidth constraint by using at
most $1$~GPU per PCIe switch.
\subsubsection{Benchmarks}
All benchmarks ran on top of a GNU/Linux~Debian~6.0.2~\textit{squeeze} with
kernel~2.6.32-5-amd64. We compiled with GCC~4.4 and linked against CUDA~5.0 and
the library ATLAS~3.9.39 (BLAS and LAPACK).
All experiments use the tile algorithms of PLASMA
\cite{ButtariLangouKurzakEtAl2009} for Cholesky (\dpotrf), LU (\dgetrf), and QR
(\dgeqrf). The QUARK API \cite{YarKhanKurzakDongarra2011} has been implemented
and extended in \xkaapi{} to support task multi-specialization: the \xkaapi{} runtime 
maintains the CPU and GPU versions for each PLASMA task.
At the task execution, our QUARK version runs the appropriate task
implementation in accordance with the worker architecture.
The GPU kernels of QR and LU are based on previous works from
\cite{AgulloAugonnetDongarraEtAl2011a,AgulloAugonnetDongarraEtAl2011} and
adapted from PLASMA CPU algorithm and MAGMA from
\cite{TomovDongarraBaboulin2010}.
%
%
Each running GPU monopolizes a CPU to manage its worker.
The remaining CPU cores are involved in the application computations.


\subsubsection{Methodology}
Each experiment has been executed at least 30 times for each set of parameters and
we report on all the figures (Fig. \ref{sec:res:dpotrf:alpha}, \ref{sec:res:potrf}, \ref{sec:res:getrf} and \ref{sec:res:geqrf}) the mean and the $95\%$ confidence interval.
The factorizations have been done in double precision floating-point operations with a PLASMA internal block (\emph{IB}) of size $128$ and tiles of size $512$.
For each of them, we plot the highest performance obtained on various matrix sizes with the discussed scheduling strategies.

In the following, DADA($\alpha$) represents DADA parametrized by $\alpha$.
We denote by DADA($\alpha$)+CP the algorithm using Communication Prediction as supplementary information.
HEFT strategy always computes the earliest finish time of a task taking into account the time to transfer data before executing the task.
\subsection{Impact of the affinity control parameter $\alpha$}
\label{sec:subsec:perf:alpha}
This section highlights the impact of the affinity control parameter $\alpha$
on the compromise between performance and data transfers. The measures have been
done with the Cholesky decomposition on matrices of size $8192 \times 8192$ and $16384 \times 16384$.
However, we present only results for the smallest size as
we observe similar behaviors for both matrix sizes.

\begin{figure*}[tb]
\centering
\subfigure[Performance of DADA($\alpha$).]{
	\includegraphics[width=0.475\textwidth]{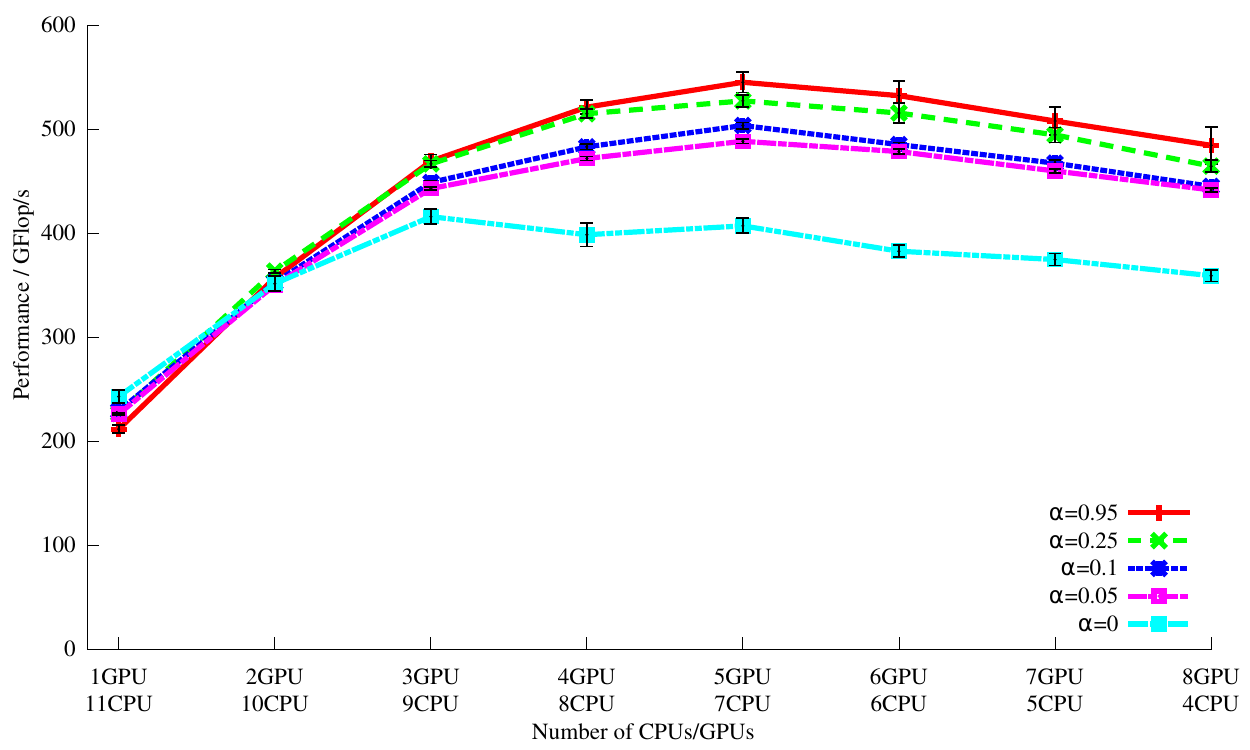}
	\label{sec:res:dpotrf:alpha:noperf:8192}
}
\subfigure[Performance of DADA($\alpha$)+CP.]{
	\includegraphics[width=0.475\textwidth]{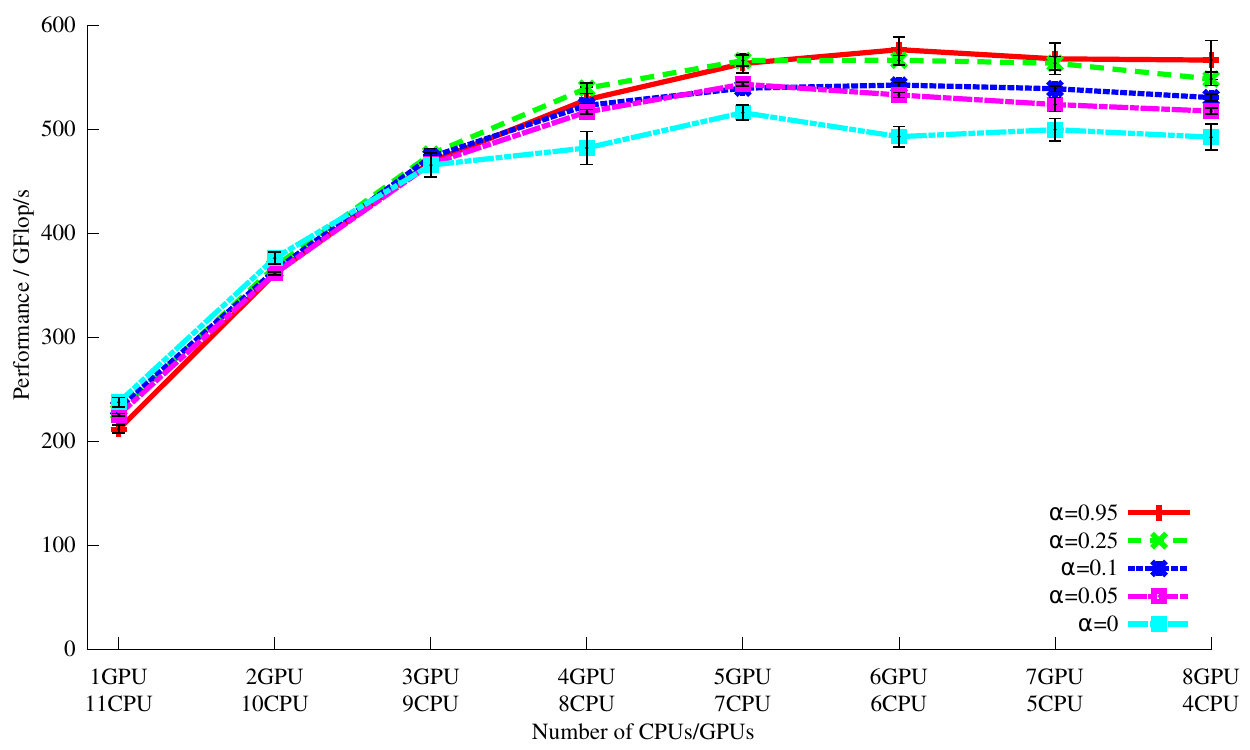}
	\label{sec:res:dpotrf:alpha:perf:8192}
}

\subfigure[Memory transfer of DADA($\alpha$).]{
	\includegraphics[width=0.475\textwidth]{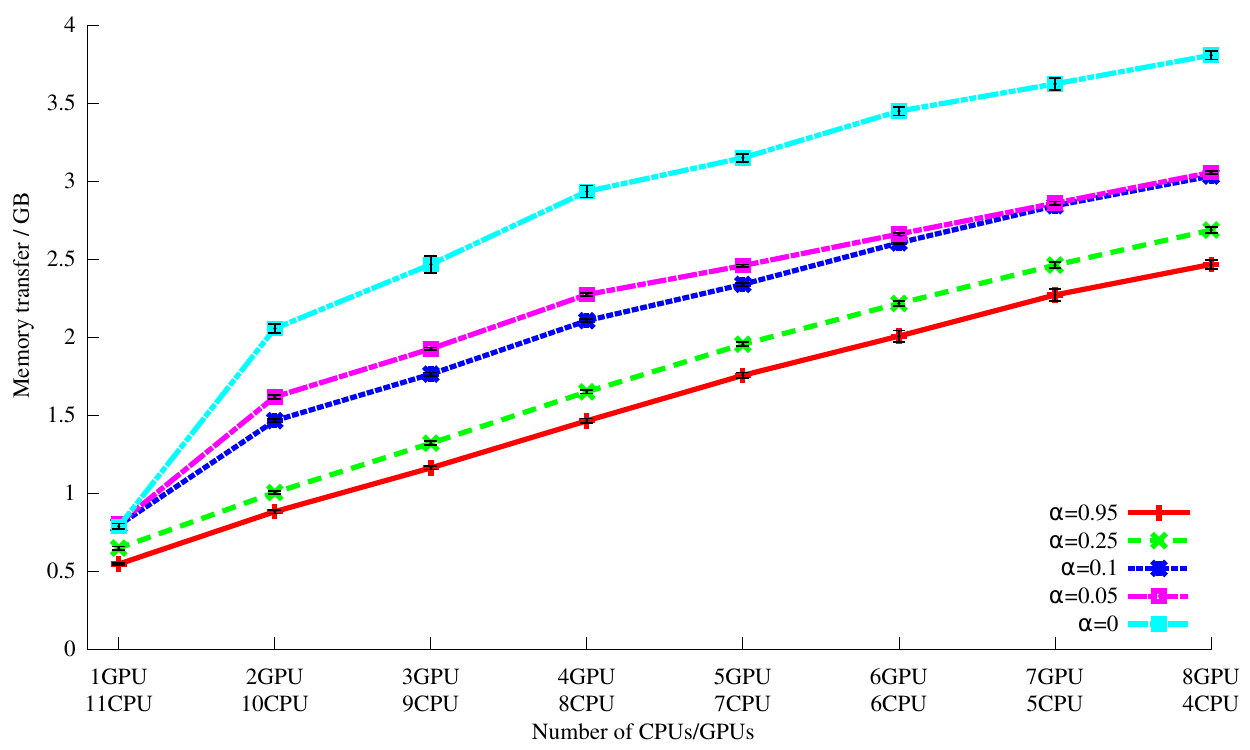}
	\label{sec:res:dpotrf:alpha:transfer:noperf:8192}
}
\subfigure[Memory transfer of DADA($\alpha$)+CP.]{
	\includegraphics[width=0.475\textwidth]{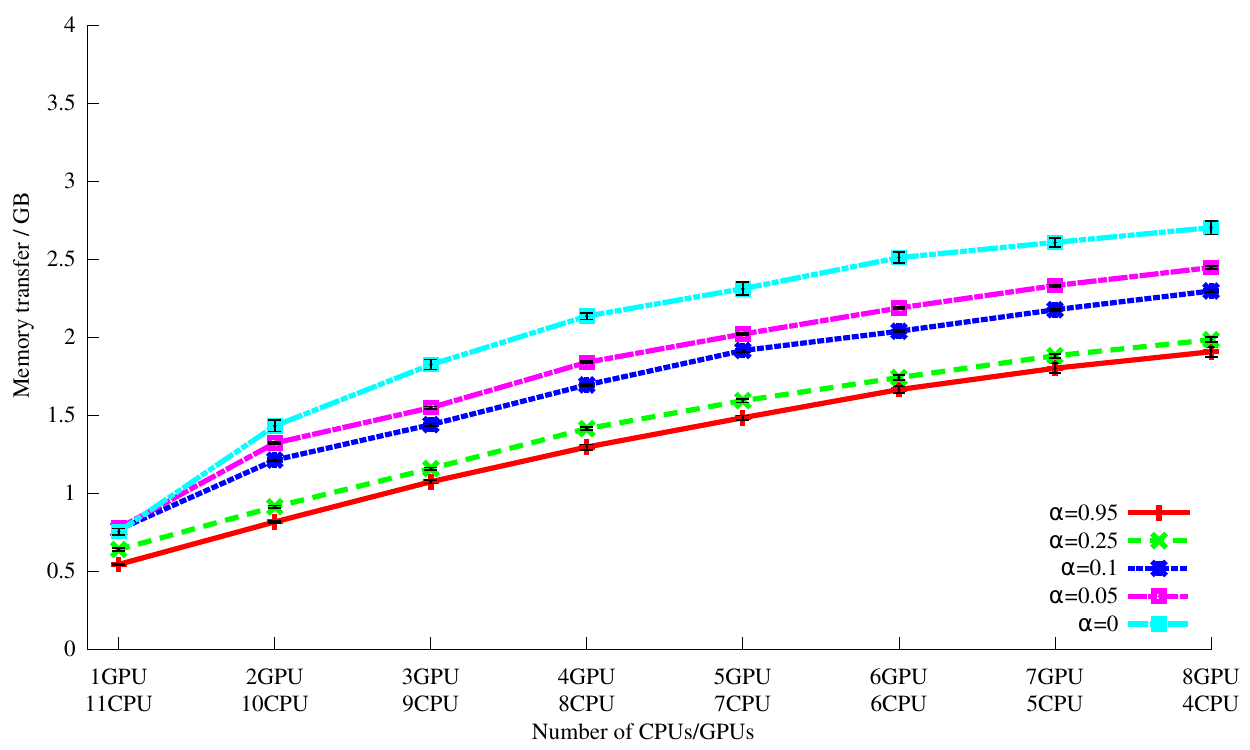}
	\label{sec:res:dpotrf:alpha:transfer:perf:8192}
}
\caption{Impact of parameter $\alpha$ on Cholesky (\dpotrf) with matrix of size $8192 \times 8192$.}
\label{sec:res:dpotrf:alpha}
\end{figure*}

Fig.~\ref{sec:res:dpotrf:alpha} shows both performance
(Fig.~\ref{sec:res:dpotrf:alpha:noperf:8192} and
\ref{sec:res:dpotrf:alpha:perf:8192}) and total memory transfers
(Fig.~\ref{sec:res:dpotrf:alpha:transfer:noperf:8192} and
\ref{sec:res:dpotrf:alpha:transfer:perf:8192}) for several values of
$\alpha$ with respect to the number of GPUs.  Both metrics
are shown without (Fig.~\ref{sec:res:dpotrf:alpha:noperf:8192} and
\ref{sec:res:dpotrf:alpha:transfer:noperf:8192}) and with
(Fig.~\ref{sec:res:dpotrf:alpha:perf:8192} and
\ref{sec:res:dpotrf:alpha:transfer:perf:8192}) communication prediction
taken into account.
Once affinity is considered (\emph{i.e.} $\alpha \ne 0$), the higher the value
of $\alpha$, the better the policy scales.
Using as little information as possible (\emph{i.e.} DADA($0$) and
no communication prediction), the policy performance does not scale with more
than two GPUs due to a too huge amount of transfers.
%
%
\subsection{Comparison of scheduling strategies}
\label{sec:subsec:perf:results}
We present in this section the results for the three kernels with matrix size
$8192 \times 8192$. Other tested sizes have the same behavior.
The idea is to evaluate the behavior
of each strategy with different work loads. Both performance and data transfers
of the policies introduced above: HEFT, DADA($0$), DADA($\alpha$) and DADA($\alpha$)+CP are studied.
\subsubsection{Experimental evaluation}
\begin{figure*}[tb]
\centering
\subfigure[Performance ($8192 \times 8192$).]{
	\includegraphics[width=0.475\textwidth]{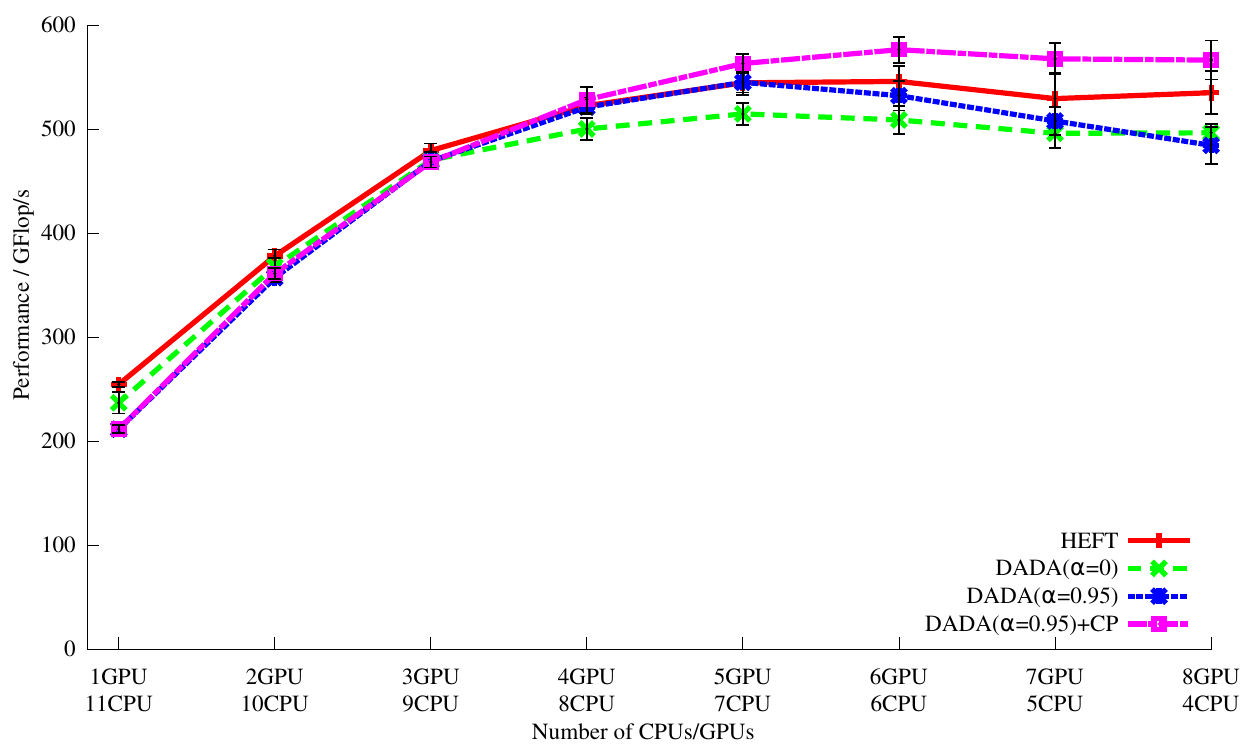}
	\label{sec:res:potrf:perf:8192}
}
\subfigure[Memory Transfer ($8192 \times 8192$).]{
	\includegraphics[width=0.475\textwidth]{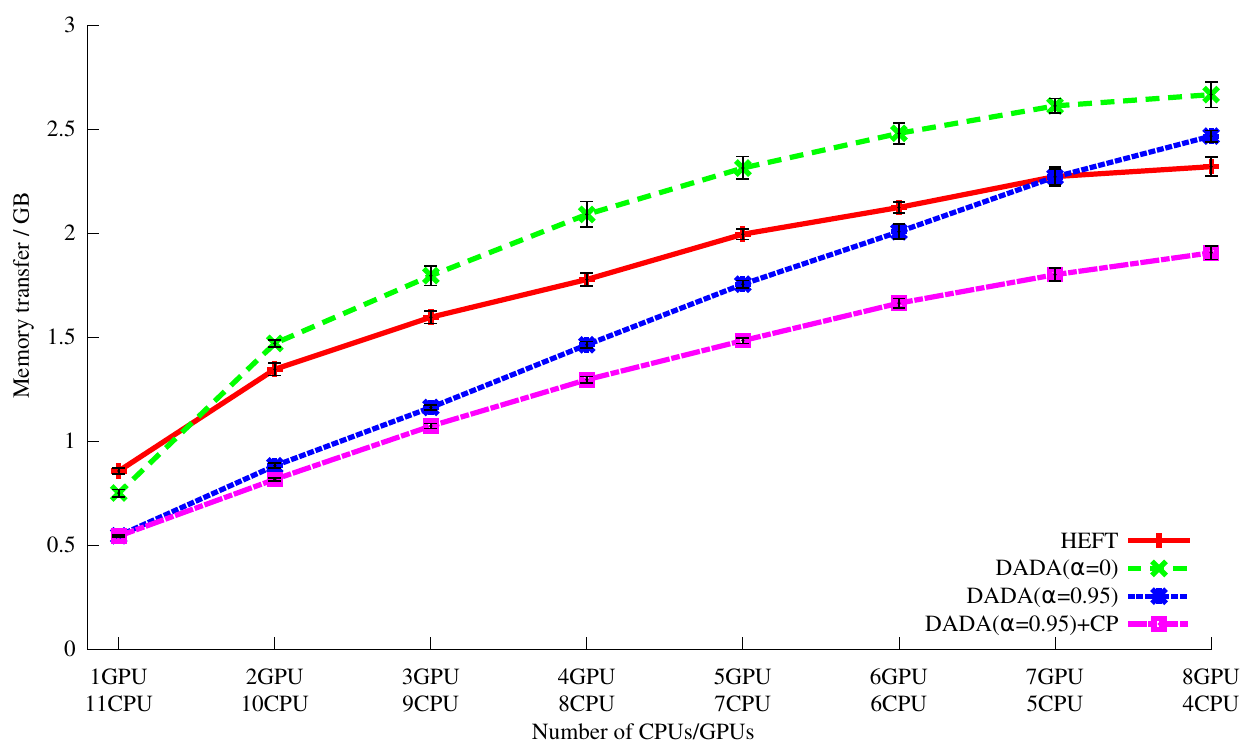}
	\label{sec:res:potrf:mem:8192}
}

%
\caption{Benchmarks of Cholesky (\dpotrf).}
\label{sec:res:potrf}
\end{figure*}

Fig.~\ref{sec:res:potrf} reports the behavior of the Cholesky decomposition
(\dpotrf) with respect to the number of GPUs used. It studies both performance
results (Fig.~\ref{sec:res:potrf:perf:8192})
 and total memory transfers (Fig.~\ref{sec:res:potrf:mem:8192}). 
All scheduling algorithms have similar performances.
DADA($\alpha$)+CP slightly better scales with the number of GPU. 
As expected DADA($\alpha$)+CP and DADA($\alpha$) are the policies with the lowest bandwidth footprint up to 6 GPU. Yet, as the number of GPU grows,
the use of communication prediction allows to reduce the communication volume with sustained high performances.

\begin{figure*}[tb]
\centering
\subfigure[Performance ($8192 \times 8192$).]{
	\includegraphics[width=0.475\textwidth]{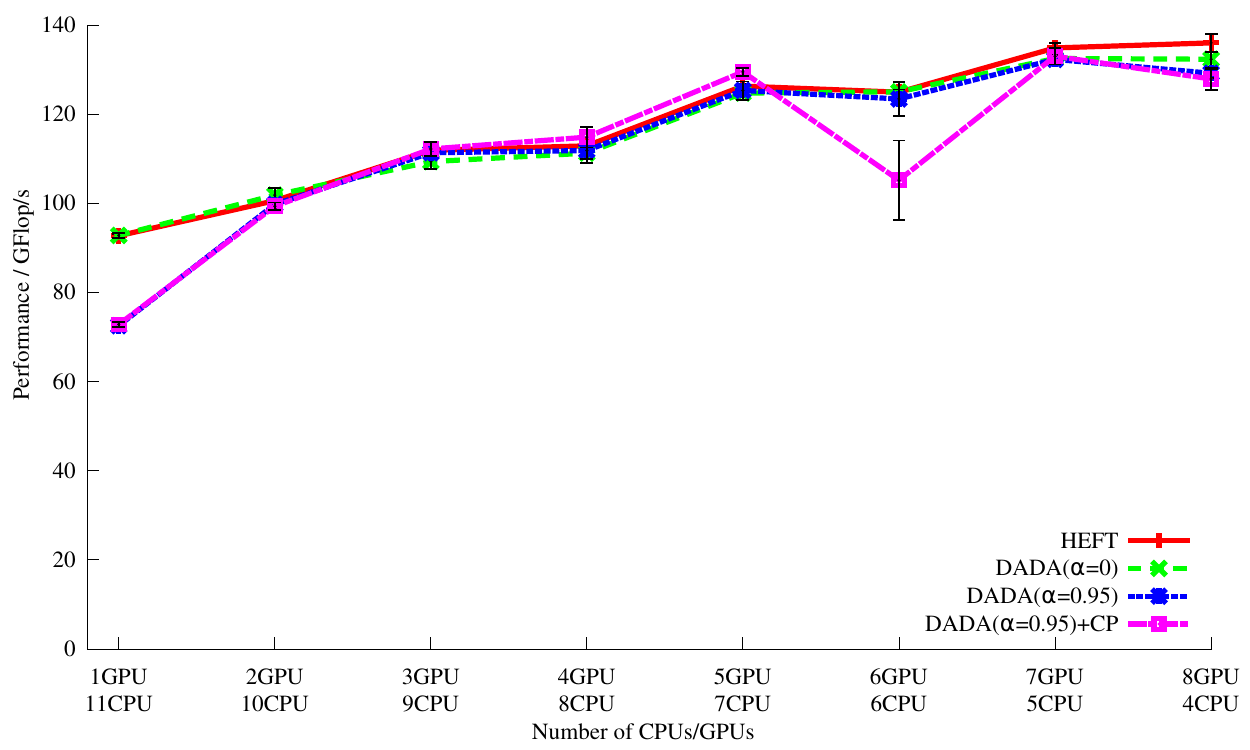}
	\label{sec:res:getrf:perf:8192}
}
\subfigure[Memory Transfer ($8192 \times 8192$).]{
	\includegraphics[width=0.475\textwidth]{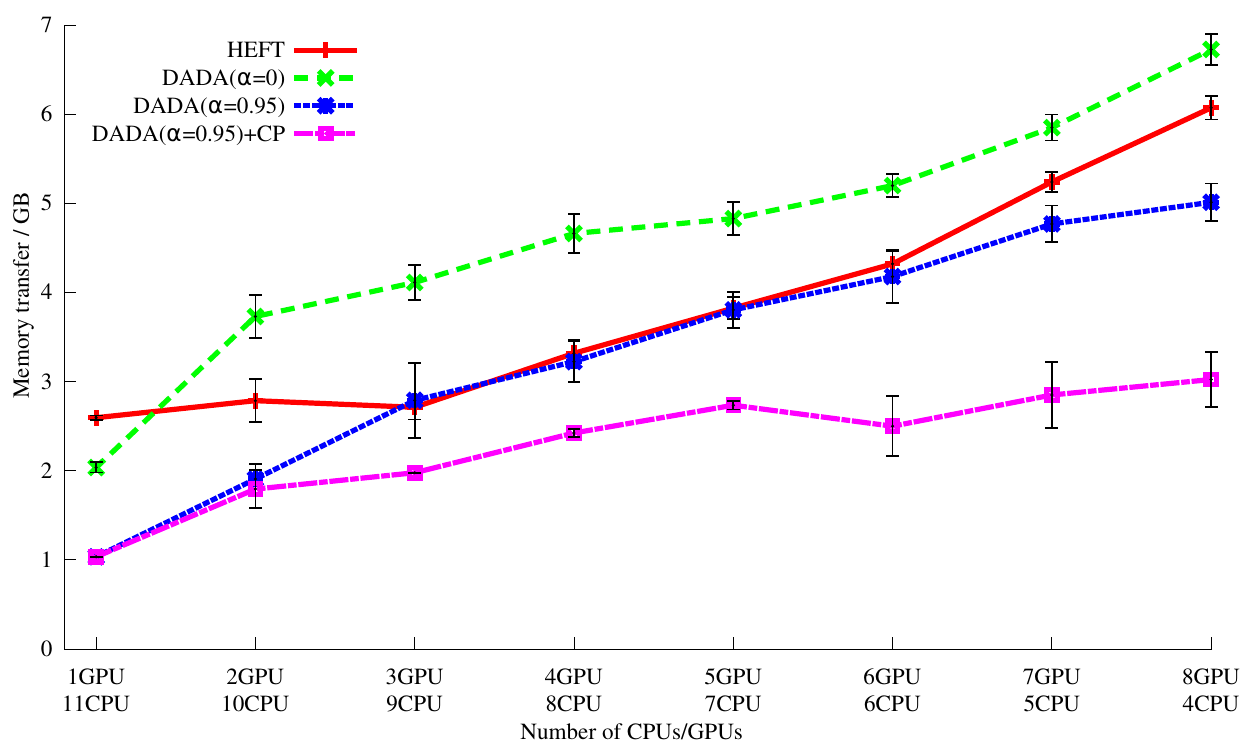}
	\label{sec:res:getrf:mem:8192}
}

%
\caption{Benchmarks of LU (\dgetrf).}
\label{sec:res:getrf}
\end{figure*}

Fig.~\ref{sec:res:getrf} reports the behavior of the LU factorization (\dgetrf).
It studies both performance results (Fig.~\ref{sec:res:getrf:perf:8192})
and total memory transfers (Fig.~\ref{sec:res:getrf:mem:8192}). 
Apart from the performance of DADA+CP for six CPUs and six GPUs (with
a large confidence interval), all scheduling policies sustain the same
performance.
Data transfers seem to have a little impact on performance. However, DADA($\alpha$)+CP generates less
memory movements than other strategies. DADA($0$) is the most costly policy while DADA($\alpha$)
and HEFT have similar impacts. 

The total memory transfers have the same shape than for the Cholesky factorization. 
Still, the gap between the curves is widening:
DADA($\alpha$)+CP is $3.5$ less demanding in bandwidth than HEFT for only a
slowdown of about $1.13$ in performance for 8 GPU.

\begin{figure*}[tb]
\centering
\subfigure[Performance ($8192 \times 8192$).]{
	\includegraphics[width=0.475\textwidth]{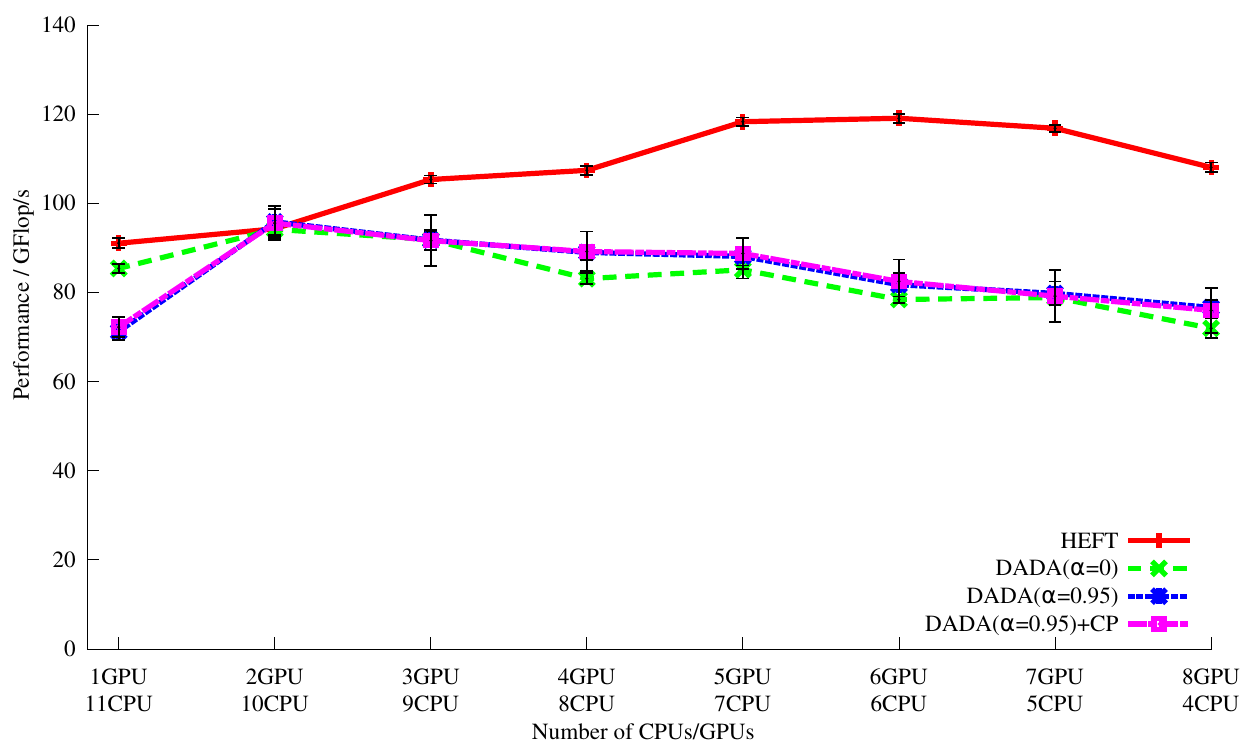}
	\label{sec:res:geqrf:perf:8192}
}
\subfigure[Memory Transfer ($8192 \times 8192$).]{
	\includegraphics[width=0.475\textwidth]{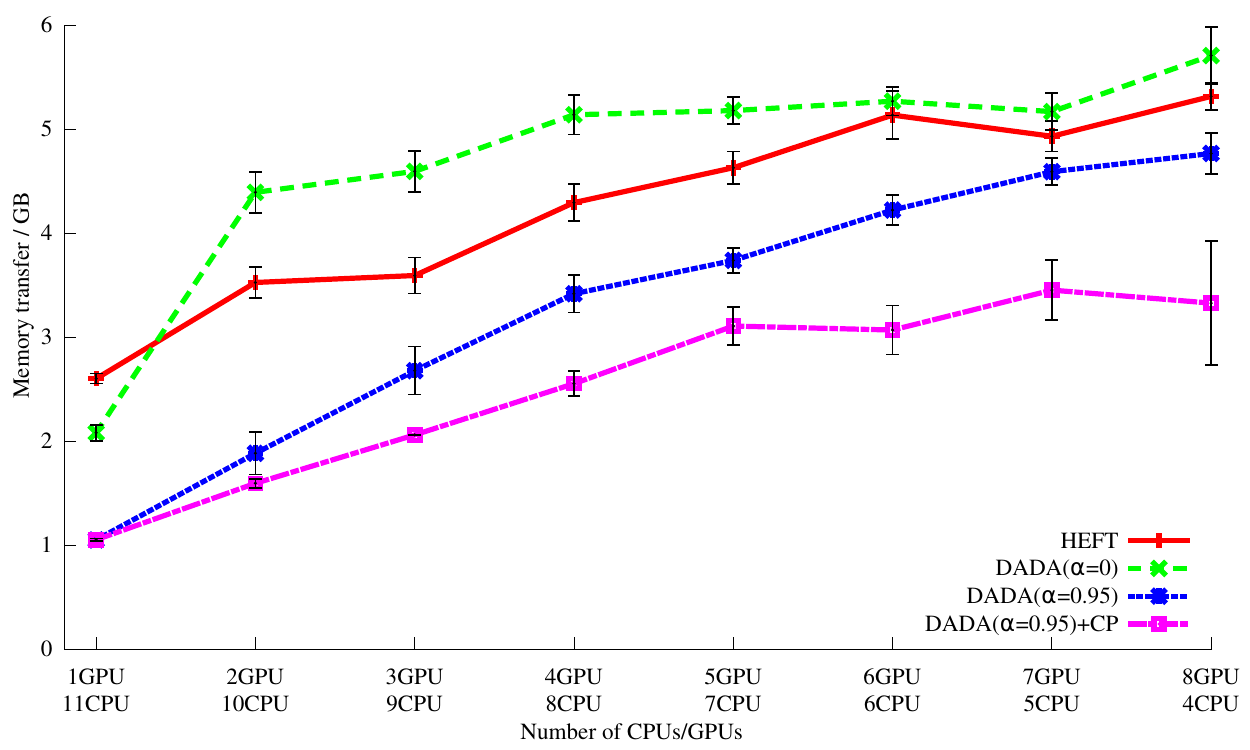}
	\label{sec:res:geqrf:mem:8192}
}

%
\caption{Benchmarks of QR (\dgeqrf).}
\label{sec:res:geqrf}
\end{figure*}

Finally, Fig.~\ref{sec:res:geqrf} reports the behavior of the QR
factorization (\dgeqrf) with respect to the number of GPUs used. Both
performance results (Fig.~\ref{sec:res:geqrf:perf:8192}-)
and total memory transfers
(Fig.~\ref{sec:res:geqrf:mem:8192}) 
are studied. All dual approximations (DADA($0$), DADA($\alpha$), DADA($\alpha$)+CP) behave the same
and are outperformed by HEFT. Even the low transfer
footprint of both DADA($\alpha$) is not able to sustain performance.
It seems that the dependencies between tasks for QR factorization have a strong impact on the schedule computed by all dual approximation algorithms.
We are still investigating this particular point.
%
%
%
%
\subsubsection{Discussion}
\label{sec:subsec:perf:discussion}
\paragraph{Communication prediction}
Affinity is a viable alternative to communication modeling. Indeed, DADA
without communication prediction is comparable to HEFT in terms of performance.
Moreover, affinity based policy combined with communication prediction allows to
reduce further more memory transfers (up to a factor $3.5$ when compared to
HEFT).
%
\paragraph{Comparison with work stealing scheduling algorithm}
For the sake of completeness, we also tested the work stealing
algorithm. However we did not plot the results in previous figures for the sake of readability. We
briefly discuss them now.
The naive work stealing algorithm is cache unfriendly, especially with small matrices as its random choices are heavily penalizing~\cite{GautierLimaMaillardEtAl2013}.
On the contrary, the affinity policies proposed here are suitable for this case.
When scheduling for medium and large matrix sizes, the impact of modeling inaccuracies grows.
Model oblivious algorithms such as work-stealing behave well by efficiently overlapping communications and computations while HEFT is induced in error by the imprecise communication prediction.
Hence, our approach is much more robust than work stealing and HEFT since it does not need a too precise communication model and adapts well to various matrix sizes.
%
%
\section{Related Works}
\label{sec:related}
%
StarPU \cite{AugonnetThibaultNamystEtAl2011}, OmpSs
\cite{BuenoPlanasDuranEtAl2012} and QUARK \cite{YarKhanKurzakDongarra2011} are
programming environments or libraries that enables to automatically schedule
tasks with data flow dependencies. OmpSs is based on OpenMP-like pragmas while
StarPU and QUARK are C libraries of function. QUARK does not schedule tasks on
multi-GPUs architecture and implements a centralized greedy list scheduling
algorithm.
OmpSs locality-aware scheduling, similar to our data-aware heuristic
from~\cite{GautierLimaMaillardEtAl2013}, computes an affinity score based on
where the data is located and its size.  Then, the task is placed on the
highest affinity resource or in a global list, otherwise. StarPU scheduler uses
the HEFT~\cite{TopcuogluHaririWu2002} algorithm to schedule all ready tasks in
accordance with the cost models for data transfer and task execution time \cite{AugonnetThibaultNamyst2010}.
Our data transfer model is based
on StarPU model with minor extension.
In the context of dense linear algebra algorithms, PLASMA
\cite{ButtariLangouKurzakEtAl2009} provides fine-grained parallel linear
algebra routines with dynamic scheduling through QUARK, which was conceived
specially for numerical algorithms on multi-CPUs architecture.
MAGMA \cite{TomovDongarraBaboulin2010} implements static scheduling for linear
algebra algorithms on heterogeneous systems composed of GPUs. Recently it has
included some methods with dynamic scheduling in multi-CPU and multi-GPU on top
of StarPU, in addition to the static multi-GPU version. In
\cite{SongDongarra2012} the authors based their Cholesky factorization on 2D
block cyclic distribution with an owner compute rule to map tasks to resources.
DAGuE \cite{BosilcaBouteillerDanalisEtAl2012} is a parallel framework focused
on multi-core clusters and supports single-GPU nodes.
Other papers reported performance results of task-based algorithms with HEFT
cost model scheduling on heterogeneous architectures for the Cholesky
\cite{AugonnetThibaultNamystEtAl2011}, LU
\cite{AgulloAugonnetDongarraEtAl2011a}, and QR
\cite{AgulloAugonnetDongarraEtAl2011} factorizations. All the results report
evaluation of single floating point arithmetics with up to $3$ GPUs. Due to the
small number of GPUs, such studies cannot observe contention and scalability.
%
\section{Conclusion}
\label{sec:conclusion}
%
We presented in this paper a new scheduling algorithm on top of the \xkaapi{} runtime.
It is based on a dual approximation scheme with affinity and has been compared to the classical HEFT algorithm
for three tile algorithms from PLASMA on an heterogeneous architecture composed of $8$~GPUs and $12$~CPUs.
Both algorithms attained significant speed up on the three dense linear algebra kernel.
Moreover, if HEFT achieves the best absolute performance with respect to DADA on QR, while DADA has
similar or better performances than HEFT on Cholesky and LU for large numbers of GPU.
Nevertheless, DADA allows to significantly reduce the data transfers with respect to HEFT.
More interesting,  thanks to its affinity criteria DADA can introduce
communication in the scheduling without too precise communication cost model
which are required in HEFT to predict the completion time of tasks.

We would like to extend the experimental evaluations on robustness of scheduling with respect
to uncertainties in cost models, especially on the communication cost which is
very sensitive to contentions that may appear at runtime.
Another interesting issue would be to study other affinity functions.
%
 \section*{Acknowledgments}
%
This work has been partially supported by the French Ministry of Defense -- DGA, the ANR 09-COSI-011-05
Project Repdyn and CAPES/Brazil.
%
\bibliographystyle{splncs03}
\bibliography{europar2014}

\begin{thebibliography}{10}
\providecommand{\url}[1]{\texttt{#1}}
\providecommand{\urlprefix}{URL }

\bibitem{AgulloAugonnetDongarraEtAl2011a}
Agullo, E., Augonnet, C., Dongarra, J., Faverge, M., Langou, J., Ltaief, H.,
  Tomov, S.: Lu factorization for accelerator-based systems. In: IEEE/ACS
  AICCSA. pp. 217--224. AICCSA '11, IEEE Computer Society, Washington, DC, USA
  (2011)

\bibitem{AgulloAugonnetDongarraEtAl2011}
Agullo, E., Augonnet, C., Dongarra, J., Faverge, M., Ltaief, H., Thibault, S.,
  Tomov, S.: {{QR} {F}actorization on a {M}ulticore {N}ode {E}nhanced with
  {M}ultiple {GPU} {A}ccelerators}. In: IEEE IPDPS. EUA (2011)

\bibitem{AugonnetThibaultNamyst2010}
Augonnet, C., Thibault, S., Namyst, R.: Automatic calibration of performance
  models on heterogeneous multicore architectures. In: Euro-Par. pp. 56--65.
  Springer-Verlag (2010)

\bibitem{AugonnetThibaultNamystEtAl2011}
Augonnet, C., Thibault, S., Namyst, R., Wacrenier, P.A.: {StarPU}: a unified
  platform for task scheduling on heterogeneous multicore architectures.
  Concurrency and Computation: Practice and Experience  23(2),  187--198 (2011)

\bibitem{BosilcaBouteillerDanalisEtAl2012}
Bosilca, G., Bouteiller, A., Danalis, A., Herault, T., Lemarinier, P.,
  Dongarra, J.: {DAGuE: A generic distributed DAG engine for High Performance
  Computing}. Parallel Computing  38(1--2),  37--51 (2012)

\bibitem{BuenoPlanasDuranEtAl2012}
Bueno, J., Planas, J., Duran, A., Badia, R.M., Martorell, X., Ayguad\'{e}, E.,
  Labarta, J.: {P}roductive {P}rogramming of {GPU} {C}lusters with {OmpSs}. In:
  IEEE IPDPS (2012)

\bibitem{ButtariLangouKurzakEtAl2009}
Buttari, A., Langou, J., Kurzak, J., Dongarra, J.: {A class of parallel tiled
  linear algebra algorithms for multicore architectures}. Parallel Computing
  35(1),  38--53 (2009)

\bibitem{GautierBesseronPigeon2007}
Gautier, T., Besseron, X., Pigeon, L.: {KAAPI}: A thread scheduling runtime
  system for data flow computations on cluster of multi-processors. In:
  PASCO'07. ACM, London, Canada (2007)

\bibitem{GautierLimaMaillardEtAl2013}
Gautier, T., Lima, J.V., Maillard, N., Raffin, B.: {XKaapi: A Runtime System
  for Data-Flow Task Programming on Heterogeneous Architectures}. In: IEEE
  IPDPS. pp. 1299--1308 (2013)

\bibitem{HermannRaffinFaureEtAl2010}
Hermann, E., Raffin, B., Faure, F., Gautier, T., Allard, J.: {M}ulti-{GPU} and
  {M}ulti-{CPU} {P}arallelization for {I}nteractive {P}hysics {S}imulations.
  In: Euro-Par. vol. 6272, pp. 235--246. Springer (2010)

\bibitem{HochbaumShmoys1987}
Hochbaum, D.S., Shmoys, D.B.: Using dual approximation algorithms for
  scheduling problems theoretical and practical results. J. ACM  34(1),
  144--162 (Jan 1987)

\bibitem{Kedad-SidhoumMonnaMounieEtAl2013}
Kedad-Sidhoum, S., Monna, F., Mouni\'e, G., Trystram, D.: Scheduling
  independent tasks on multi-cores with gpu accelerators. In: 11th HeteroPar
  Workshop (2013)

\bibitem{LimaGautierMaillardEtAl2012}
Lima, J.V.F., Gautier, T., Maillard, N., Danjean, V.: {E}xploiting {C}oncurrent
  {GPU} {O}perations for {E}fficient {W}ork {S}tealing on {M}ulti-{GPUs}. In:
  24th SBAC-PAD. pp. 75--82. IEEE, New York, USA (2012)

\bibitem{SongDongarra2012}
Song, F., Dongarra, J.: {A} scalable framework for heterogeneous {GPU}-based
  clusters. In: ACM SPAA. pp. 91--100. ACM, New York, NY, USA (2012)

\bibitem{TomovDongarraBaboulin2010}
Tomov, S., Dongarra, J., Baboulin, M.: {Towards dense linear algebra for hybrid
  GPU accelerated manycore systems}. Parallel Computing  36(5-6),  232--240
  (2010)

\bibitem{TopcuogluHaririWu2002}
Topcuoglu, H., Hariri, S., Wu, M.Y.: Performance-effective and low-complexity
  task scheduling for heterogeneous computing. IEEE TPDC  13(3),  260--274
  (2002)

\bibitem{YarKhanKurzakDongarra2011}
YarKhan, A., Kurzak, J., Dongarra, J.: Quark users' guide: Queueing and runtime
  for kernels. Tech. Rep. ICL-UT-11-02, University of Tennessee (2011)

\end{thebibliography}
%
\end{document}